\documentstyle[12pt]{article}
\normalbaselines
\def\beq{\begin{equation}}
\def\eeq{\end{equation}}

\title{\LARGE \bf  INTEGRALS OF MOTION\\ FOR KICKED QUANTUM SYSTEMS\\
INFN-NA-IV-93/48~~~~~~~~~~~~~~~~~~~~~~~~~~~~~~~~~~~~~~~~~DSF-T-93/48
\\ [7mm] }

\author{ V.~I.~Man'ko\\
{\it Lebedev Physics Institute, Moscow, Russia}\\[3mm]}

\begin{document}
\maketitle
\newpage
Abstract:

The generalised quasienergy states are introduced as eigenstates of
the new integral of motion for periodically and nonperiodically
kicked quantum systems.The photon distribution function of
polymode generalised correlated light expressed in terms of
multivariable Hermite polynomials is discussed and the relation
of its properties to Schrodinger uncertainty relation is given.

(Invited talk at Symposium on the Foundation of Modern Physics,
K\"oln, June 1993; to appear in Proceedings of the Symposium,
World Scientific Publishers)

\newpage
\section{Introduction}

The nonstationary guantum systems have specific integrals of motion
which depend on time in Scrodinger representation (see, for example
[1]).
The integrals of the motion which are  the operators
linear in position and momentum and which depend on time in
the Schrodinger
representation have been constructed for the forced oscillator with
time-dependent frequency and for the charged particle moving in time
dependent magnetic field in [2] and [3].The linear
integrals of the motion for the nonstationary systems with Hamiltonian
which is a generic multidimensional quadratic form in positions and
momenta operators have been described,for example,in [4].
In all these systems
the dependence of
the parameters on time was the arbitrary one including
both very slow
adiabatic case and very fast instant change of parameters.

The important partial case of nonstationary systems is the
kicked systems on which the external forces act during very short
time in comparison with all the periods of possible vibrations in the
system under study.The aim of talk is to discuss properties of the
time-dependent invariants for the kicked systems because several
interesting
phenomena
such as extension of Casimir effect to nonstationary conditions
or deformations of particle distribution functions
may be demonstrated for the kicked systems.
The Casimir
effect is the phenomenon of attraction of two neutral plates between
which is the vacuum of the electromagnetic field due to
the dependence of the vacuum energy on the distance between the plates.
If one moves the plates the Casimir forces produce
the work and due to conservation of the energy the vacuum state
becomes the state with photons.So we have nonstationary Casimir effect
which is the phenomenon of generation of photons from the vacuum.

The electromagnetic field may be considered as the sum of modes and each
mode being the oscillator with its own frequency.
When the plates move these oscillators are just the oscillators with
time dependent frequencies.So we treat the nonstationary Casimir
effect as the effect existing for any system with time dependent
parameters.In this case the frequences are these parameters
(like distance between the plates).Thus the linear invariants
found in [2] and [4] for one dimensional and
multidimensional oscillators are in the case of nonstationary
Casimir effect the integrals of
motion for electromagnetic field with time varying boundaries.

Quantum systems with energy spectra are the systems
with stationary
Hamiltonians.The dynamics of these systems is described by the transitions
between the energy levels.The nonstationary quantum systems
have no energy levels.But for periodical quantum systems
the notion of quasienergy
levels has been introduced in [5] and [6].The main point
of the quasienergy concept is to relate the quasienrgies to the eigenvalues of
the Floquet operator which is equal to the evolution operator of a quantum
system taken
at a given time moment.

In [7] the connection of Floquet operator and quasienergies with
time-dependent integrals of motion has been found.
Following this article we will discuss the relation of the Floquet
operator to integrals of motion and  introduce a new operator which is
the integral of motion and has the same quasienergy spectrum that the Floquet
operator has.Implicitly this result was contained in [4].

Thus,the quasienergy of the periodically
kicked quantum systems may be shown to be the time-dependent integral
of motion for such systems.If the system is kicked nonperiodically it is
possible nevertherless to find out the time-dependent integrals of motion
and to introduce the generalized quasienergy states [7].
It gives the
possibility to extend the analisys of quantum chaos phenomenon which is
usually based on the studying the quasienergy spectra of the kicked
systems to the
case of nonperiodically kicked systems.

Another physical phenomenon related to the kicking is the
creating the photons in squeezed
states from vacuum due to the nonstationary Casimir effect mentioned above.
We will discuss how
the electromagnetic field (or pion or boson field of any kind)
changes their statistical properties
due to influence of the kicking.As an example the state with the Wigner
function of generic Gaussian form will be considered as the final result
of kicking.

\section{Generalized quasienergies}

Following [7] we will discuss the properties of
quasienergies and generalized quasienergies for periodically and
nonperiodically kicked quantum systems.

If one has the system with hermitian Hamiltonian $H(t)$ such that
$H(t+T)=H(t)$ the unitary evolution operator $U(t)$ is defined as follows
\beq
|\psi,t>=U(t)|\psi,0>
\eeq
where $|\psi,0>$ is a state vector of the system taken at the initial
time moment.Then by definition the operator $U(T)$ is called the Floquet
operator and its eigenvalues have the form
\beq
f=exp(-iET)
\eeq
where $E$ is called the quasienergy and the corresponding eigenvector is
called the quasienergy state vector.The spectra of quasienergy may be either
discrete or continuous ones (or mixed) for different quantum systems.
For multidimensional
systems with quadratic Hamiltonians the quasienergy spectra have been related
to real symplectic group $ISp(2N,R)$ and found
explicitly in [4].

We want
to answer the following questions.
Is the Floquet operator $U(T)$ the
integral of motion?The operator $U(T)$ does not satisfy the relation
\beq
dI(t)/dt+i[H(t),I(t)]=0,(\hbar=1)
\eeq
which defines the integral of motion $I(t)$.So,the Floquet operator $U(T)$
is not the integral of motion for the periodical nonstationary quantum
systems.But as it was pointed out in [4] any operator of the
form
\beq
I(t)=U(t)I(0)U^{-1}(t)
\eeq
satisfies the equation (3) and this operator is the integral of motion for
the quantum system under study.Let us apply this ansatz to the case of
periodical quantum systems.We introduce the unitary operator $M(t)$ which
has the form
\beq
M(t)=U(t)U(T)U^{-1}(t).
\eeq
This operator is the integral of motion due to the construction given
by the formula (4) for any integral of motion.The spectrum of the new invariant
operator $M(t)$ coincides with the spectrum of the Floquet operator $U(T)$.We
have proved that since quasienergies are defined as eigenvalues of the
integral of motion $M(t)$ they are conserved quantities.\\The given
construction permits us to introduce new invariant labels for nonperiodical
systems,for example, with the time-dependence of the Hamiltonian
corresponding to quasicrystal structure in time described by two (or more)
characteristic times.For such systems the analog of the invariant Floquet
operator (5) will be given by the relation
\beq
M_{1}(t)=U(t)U(t_{1})U(t_{2})U^{-1}(t).
\eeq
The eigenvalues of the operator $M_{1}(t)$  are the consrved
quantities and they characteryse the nonperiodical quantum systems
in the same manner as quasienergies describe
the states of periodical quantum systems.
This operator determines the generalized
quasienergies and its obvious analog describes
the geometrical phase which is the characteristic of special
nonperiodical
system for which the
parameters of the Hamiltonian take their initial values
after some time $T$ [8].

The quasienergy spectrum of periodically kicked quantum systems may
be connected with quantum chaos phenomenon (see [9],
[10]).In [11] the integral of motion
for delta-kicked nonlinear oscillator has been found to exist even
in the case of chaotic behaviour.In [12] the symmetry group
criterium for the periodically delta-kicked systems has been found to
obtain either regular or chaotic behaviour of these systems.The criterium
relates the Floquet operator spectrum to the conjugacy classes
of the system
symmetry group.The results of [12]
may be applied to the nonperiodical systems too.
The generalized quasienergy spectrum is determined by the conjugacy class
of the same group to which belongs the integral of motion (6).For quadratic
systems it is the same real symplectic group $Sp(2N,R)$.In the case of two
characteristic times the classification of the possible either chaotic or
regular regimes of the system under study coincides formally
with the classification
for the delta-kicked periodical quantum systems given in
[12].

The geometrical phase
is defined as the phase of eigenvalue of the evolution operator $U(T)$ where
$T$ is the time moment at which the parameters of Hamiltonian take their
initial values.Using similar arguments we can answer
the same question as for the quasienergies.Is
the geometrical phase the integral of motion of the nonstationary and
nonperiodical quantum system?The answer is "yes" because such nonperiodical
system has characteristic time $T$.
Due to this the operator $U(t)U(T)U^{-1}(t)$
is the integral of motion of the system.

{}From the point of view of given analisys the nonperiodical systems with
several characteristic times
have to demonstrate similar physical properties
that are usually considered as properties of purely periodical quantum
systems.Thus,the types of chaotic and regular regimes of the kicked
quantum systems have to be the
same for both periodical and nonperiodiacal
kicks and group classification of regular and irregular behaviour of
kicked systems given in [12] may be easily extended to
the case of nonperiodically kicked systems.

\section{Photon distributions}

Let us consider as nonstationary quantum system the photons
in a resonator.We want to discuss the photon distribution function
which is the distribution for the generalized correlated state
found in [13].Below we follow this work.
Due to the varying boundaries and time dependence of the refraction
index of
the media in a resonator the mixed squeezed state of the $N$-mode
light with density operator $\hat \rho $ may emerge.It is
described by Wigner
function $W({\bf p},{\bf q})$ of the generic Gaussian form
which contains $2N^2+3N$ real parameters.$2N$ parameters are mean
values of light quadratures $<{\bf p}>$ and quadratures
$<{\bf q}>$  and other $2N^2+N$ parameters are
matrix elements of the real symmetric dispersion
matrix $m$ with four $N$-dimensional block matrices three of which are
\begin {eqnarray}
m_{11}=\sigma _{\bf p},\\
m_{12}=\sigma _{\bf p\bf q},\\
m_{22}=\sigma _{\bf q}.
\end {eqnarray}
Below we will use the invariant parameters
\beq
T=Tr~{m}
\eeq
and
\beq
d=\det m.
\eeq
Also we will use the other invariant coefficients of the polynomial
\beq
P(x)=\det (m-x\bf 1)
\eeq
where $\bf 1$ is $2N$-dimensional identity matrix.

We will introduce the notations
\beq
{\bf Q}=({\bf {p}},{\bf {q}})
\eeq
where $2N$-dimensional vector $\bf Q$ consists of $N$ components of light
quadratures
$p_{1},...,p_{N}$ and $N$ components of quadratures $q_{1},...,q_{N}$.
The generic gaussian Wigner function has the form
(see,for example,[1])
\beq
W({\bf {p}},{\bf {q}})
=d^{-\frac{1}{2}}\exp [-(2)^{-1}[({\bf {Q-<Q>}})m^{-1}({\bf {Q-<Q>}})].
\eeq
The parameters $<\bf p>$ and $<\bf q>$ are given by the formulae
\begin {eqnarray}
<{\bf p}>=Tr~ \hat \rho \hat {\bf p} ,\\
<{\bf q}>=Tr~ \hat \rho {\hat {\bf q}},
\end {eqnarray}
where the operators $\hat {\bf p}$ and $\hat {\bf q}$
are the quadrature components of
photon creation $\bf a\dag $ and the annihilation ${\bf a}$ operators
\begin {eqnarray}
\hat {\bf p}=\frac{{\bf a}-{\bf a\dag}}{i\sqrt 2},\\
\hat {\bf q}=\frac{{\bf a}+{\bf a\dag}}{\sqrt 2}.
\end {eqnarray}
Due to the physical meaning of the dispersions the diagonnal elements
of the matrix $m$  must be nonnegative numbers,so the invariant parameter
$T$ Eq.(10) is a positive number.Also the determinant $d$ Eq.(11)
of the dispersion
matrix must be positive.In fact the matrix $m$ must be
positive-definite.

To obtain the photon distribution function we have to calculate the probability
$P _{\bf n}$ to have $\bf n$
photons in the state with the density operator $\hat \rho$.
Here the vector $\bf n$ has $N$ components $n_i$ which are nonnegative
integers.
This probability is given by the formula
\beq
P_ {\bf n}=Tr~ {\hat \rho}|{\bf n}><{\bf n}|,\\n_{i}=0,1,2,...~~~i=1,2,...,N
\eeq
where the number states $|{\bf n}>$ are the eigenstates of the number operator
$\bf a\dag a$
\beq
{\bf a\dag a|n>=n|n>}.
\eeq
The function $P_ {\bf n}$ may
be obtained if one calculates the generating function
for the matrix
elements $\rho _{\bf mn}$ of the density operator $\hat \rho$ in
the Fock basis.
This generating function is the matrix element of the density
operator in the coherent state basis
\beq
< {\bf\beta} | \hat \rho | {\bf\alpha} > =
\exp ( \frac {-|{\bf\alpha}|^2}{2} - \frac {|{\bf\beta}|^2}{2} )
 \sum_{{\bf m,n=0}}^{\infty }
\frac{ ({\bf \beta }^{*})^{{\bf m}} {\bf \alpha}^{\bf n}}
{   ( {\bf m!n!} )^{\frac {1}{2} }   }
\rho_{{\bf mn}}
\eeq
Here and below we use notations:$\alpha $and $\beta $ are $N$-dimensional
vectors with complex components and
\begin {eqnarray}
{\bf n}!=n_{1}!n_{2}!...n_{N}!,\\
{\bf \alpha }^{\bf n}=\alpha _{1}^{n_1}\alpha _{2}^{n_2}...\alpha _{N}^{n_N}
\end {eqnarray}
and
\beq
\sum_{{\bf m,n=0}}^{\infty}=\sum_{m_{1}=0}^{\infty}...\sum_{m_{N}=0}^{\infty}
\sum_{n_{1}=0}^{\infty}...\sum_{n_{N}=0}^{\infty}.
\eeq
We have
\beq
P_ {\bf n}=\rho _{{\bf nn}}.
\eeq
The $N$-mode coherent
state $|{\bf \alpha}>$ is the normalized eigenstate of the
annihilation operator
\beq
{\bf a}|{\bf \alpha}>={\bf \alpha}|{\bf \alpha}>.
\eeq
In terms of Wigner function the density operator in the coherent state
representation has the form of $2N$-dimensional overlap integral [1]
\beq
<{\bf \beta}|\hat \rho|{\bf\alpha}>=
\frac {1}{({2\pi })^{N}}\int W({\bf p},{\bf q})W_{{\bf\alpha}{\bf \beta} }
({\bf p},{\bf q})d{\bf p}d{\bf q},
\eeq
where the function $W_{{\bf \alpha} {\bf\beta }}({\bf p},{\bf q})$
is the Wigner function of the
operator $|{\bf\alpha}><{\bf\beta}|$.It has the form [1]
\beq
W_{{\bf\alpha} {\bf\beta }}({\bf p},{\bf q})=2^{N}
\exp [-\frac {|{\bf\alpha} |^2}{2}-\frac {|{\bf\beta}|^2}{2}-
{\bf\alpha} {{\bf\beta}}^{*}-{\bf p}^{2}-{\bf q}^{2}+
\sqrt 2{\bf\alpha} ({\bf q}-i{\bf p})+\sqrt 2{{\bf\beta}}^{*}
({\bf q}+i{\bf p})].
\eeq
Let us introduce the $2N$-dimensional vector
\beq
{\bf \gamma}=({\bf \beta }^{*},{\bf \alpha })
\eeq
which is composed from two $N$-dimensional vectors.Let us introduce also
$2N$- dimensional unitary matrix $u$ which has four $N$-dimensional block
matrices
\beq
u_{11}=-u_{12}=-iu_{21}=-iu_{22}=-\frac {i}{\sqrt 2}{\bf 1}
\eeq
and ${\bf 1}$ is $N$-dimensional identity matrix.

Since the integral given by Eq.(27) is the Gaussian one it may be easily
calculated.
So,we have
\beq
<{\bf \beta}|\hat \rho|{\bf\alpha}>=P_{0}
\exp (-\frac{|{\bf \gamma }|^2}{2} )
\exp [-\frac{1}{2}{\bf \gamma }R{\bf \gamma }+{\bf \gamma }R{\bf y}]
\eeq
where the symmetric $2N$-dimensinal matrix $R$
has the matrix elements expressed in terms
of the dispersion matrix $m$ as follows
\beq
R=-2u^{-1}(\frac {m^{-1}}{2}+1)^{-1}u^{*}+\Sigma _{x}
\eeq
The matrix $ \Sigma _{x}$ is $2N$-dimensional analog of Pauli matrix.
The $2N$-dimensional vector ${\bf y}$ is given by the relation
\beq
{\bf y}=2u^{t}(1-2m)^{-1}<{\bf Q}>
\eeq
where the matrix $u^t$ is transposed one and the factor $P_{0}$ has the form
\beq
P_{0}=[det (m+\frac {1}{2})]^{-\frac {1}{2}}
\exp[-<{\bf Q}>(2m+1)^{-1}<{\bf Q}>]
\eeq

We obtain for the photon distribution function $P_{{\bf }n}$ the expression
\beq
P_{{\bf n}}=P_{0}\frac{H_{{\bf nn}}^{\{R\}}({\bf y})}{{\bf n}!}.
\eeq
Here the matrix $R$ determining the Hermite polynomial is given by the
formulae (32) and  argument of Hermite polynomial is given by the
expression (33).The expression (35) is the partial case of the matrix
elements of the density operator in Fock states basis obtained in [14]
by canonical transform method.

The behaviour of the distributions may be wavy function of photon
numbers as it is in one-mode case [15],[16].

Mean values of photons in each
mode have the form [13]
\beq
<n_{j}>=\frac {1}{2}(\sigma _{p_{j}p_{j}}+\sigma _{q_{j}q_{j}}-1)+
\frac {1}{2}(<p_{j}>^{2}+<q_{j}^{2}).
\eeq
Photon number variances are given by
expression
\beq
\sigma _{n_{j}}=\frac {1}{2}(T_{j}^{2}-2d_{j}-\frac {1}{2})+
<Q_{j}>m_{j}<Q_{j}>
\eeq
where $T_j$ and $d_j$ are the trace and the determinant of
the photon quadrature $2$x$2$-dispersion matrix  $m_j$ of the j-th mode only
and the 2-vector $Q_{j}$ has components $p_{j},q_{j}$.
The correlations of photon numbers in different modes may be expressed
analogously.

\section{Uncertainty relation and distributions}

Following [17]
for one-mode case we
consider the photon distribution function (35)
expressed in terms of the series of the products of two Hermite
polynomials.

There exists the expression of the Hermite polynomial of two
variables with equal indicies in terms of the products of two
Hermite polynomials of one variable[14].
Using this formula one obtains
from the general expression (35) the following photon distribution
function
\beq
P_{n}=P_{0}n!
\sum_{k=0}^{n}(\frac {R_{11}R_{22}}{4})^{n/2}
(-\frac {2R_{12}}
{\sqrt {R_{11}R_{22}}})^{k}(n-k)!^{-2}k!^{-1}
|H_{n-k}(\frac {R_{11}y_{1}+R_{12}y_{2}}{\sqrt {2R_{11}}})|^{2}.
\eeq
Here the factor $P_{0}$ is given by the formula (34),the matrix elements of
the matrix $R$ are given by the formulae (32) and the numbers $y$
are given by the formula (33).If the state is the pure squeezed and
correlated state the determinant of the quadrature dispersion matrix is
equal to $\frac{1}{4}$ and the matrix element $R_{12}=0$ due to the formula
(32).In this case only the first term in the series is not equal to zero
and this term gives the known expression
for the photon distribution function
of the pure and correlated state obtained by another method in [16].
The above
expression  is convinient to discuss the connection of the uncertainty
relations with the quantum distribution functions.In fact the formula
for the Gaussian Wigner function (14) formally coincides with the classical
Gaussian density of probability in the particle phase space.Then we have
to answer the question in what aspects these formulae must be considered
as essentially different ones.By intuition it is clear that the uncertainty
relation must distinguish the classical and quantum Gaussian densities
of probabilities having absolutely the same form (14).
And in fact in addition
to the usual restriction for the dispersion matrix for the Gaussian
classical distribution function in the phase space which is the condition
of nonnegativity of this dispersion matrix the quantum mechanics demands
the inequality for the determinant of this matrix $d>1/4$.This inequality
is the Schrodinger uncertainty relation.The formula (38) permits
to relate this inequality with physically obvious condition that
probability to find $n$ photons must be nonnegative because
the determinant
of the dispersion matrix is the parameter on which depends the photon
distribution function.We see that all the terms in the expression (38)
are obviously positive ones except the term containing
the number $-2R_{12}$ which may change the sign for odd powers $k$ if
it is not positive itself.It means that for
natural condition of positiveness of photon distribution function it
is nesessary to have inequality
\beq
R_{12}<0
\eeq
But this inequlity is equivalent to the mentioned above
inequality for the dispersion
matrix determinant $d$ which implies the uncertainty relation.Thus
existence of the connection of the Wigner function (14) with the photon
distribution funtion (38) is consistent only if the uncertainty relation
holds.So we clarify the mechanism how uncertainty relation in
phase space of electromagnetic field oscillator
influences the form of photon distribution function.

As we discussed the
nonstationary Casimir effect produces the changes in statistical
properties of the photons.For example,it changes Poisson distribution
of photons to become the discussed distribution expressed
in terms of multivariable Hermite polynomials.
Also it deforms  Planck distribution to become [14]
\begin{equation}
\bar n = \frac{1}{e^{\frac{\hbar \omega }{kT}}-1} +|v|^{2}\coth\frac
{\hbar \omega }{2kT}+|\delta |^{2},
\end{equation}
where the correction to usual Planck distribution term contains parameters
$|v|^{2}$ and $|\delta |^{2}$.These parameters depend on the characteristics
of kicking [1],[14].So,a generic kicking produces from
ground state the squeezed mixed state with deformed Planck distribution and
the number state distribution function which is expressed expliciltly in terms
of Hermite polynomials of several variables.

It is interesting to note that if to take into account a possible
nonlinearity of classical electrodynamics we can have another reason
for deforming photon distribution formulae.
In [18]
it was suggested that the possible nonlinearity of the electromagnetic
field vibrations which must exist for very large amplitudes
corresponding
to high densities of the field energy may be considered as the
nonlinearity
described by the q-oscillator. It was seen that it is subject to
 nonlinear vibrations with a special kind of
dependence of the frequency on the amplitude and the influence of such
nonlinearity on Bose distribution
function was evaluated.
So,for the thermal state we have the deformed
Planck distribution formula [18] of the form
\beq
(\bar n)_{q} \simeq \frac {1}{e^{\frac{\hbar \omega}{kT}}-1} - \kappa^2
\frac{\hbar \omega}{kT}\frac{e^{\frac{3\hbar\omega}{kT}} + 4 e^{\frac{2\hbar
\omega}{kT}} + e^{\frac{\hbar \omega}{kT}}}{(e^{\frac{\hbar \omega}{kT}} - 1)^4
}.
\eeq
Here the first term is the usual Planck distribution formula and its
correction is proportional to the square of the nonlinearity parameter
$\kappa$.\\
It means that black body radiation formula changes due to the nonlinearity
of the electromagnetic field vibrations.So nonstationary Casimir effect
and q-nonlinearity of electromagnetic field vibrations produce deformations
of Planck distribution formula.But the temperature dependence of
corrections to Planck formula is different and it gives a possibility
to distinguish the influence of these effects.

\begin {center}
{\LARGE\bf References}\\
\end {center}

[1].V.V.Dodonov and V.I.Man'ko "Invariants and evolution of
nonstationary quantum systems",Proceedings of Lebedev Physical institute
{\bf 183},ed.by M.A.Markov,Nova Science Publishers,
Commack,N.Y.(1989).

[2].I.A.Malkin,V.I.Man'ko and D.A.Trifonov,Phys.
Lett.{\bf 30A},414(1969).

[3].I.A.Malkin and V.I.Man'ko,Phys.Lett.{\bf 32A},243(1970).

[4].I.A.Malkin and V.I.Manko "Dynamical symmetries and coherent
states of quantum systems"(in Russian),Nauka Publishers,Moscow(1979).

[5].Ya.B.Zeldovich,Sov.Phys.JETP.{\bf 24},1006(1967).

[6].V.I.Ritus,ibid,p.1041.

[7].V.I.Manko,Proc.of 2nd International Workshop on Squeezed states
and Uncertainty Relations,Moscow,Russia,May 25-29,1992,ed.D.Han,
Y.S.Kim and V.I.Manko,p.405,NASA(1993).

[8].M.V.Berry,Proc.Roy.Soc.London {\bf 392},45(1984).

[9].F.Haake,"Quantum signature of chaos",Springer-Verlag,Berlin
(1991).

[10].B.V.Chirikov,Phys.Rep.{\bf 52},263(1979).

[11].V.I.Man'ko and F.Haake,Ann.Physik {\bf 1},302(1992).

[12].G.Karner,V.I.Man'ko and L.Streit,Repts.Math.Phys.{\bf
29},177(1991).

[13].V.V.Dodonov,O.V.Man'ko and V.I.Man'ko,Multivariable
Hermite polynomials and photon distribution ,University of Napoli
preprint
INFN-NA-IV-93/36,DSF-T-93/36(1993).

[14].V.V.Dodonov,V.I.Man'ko and V.V.Semjonov,
Nuovo Cimento B {\bf 83},145(1984).

[15] W.Schleich and J.A.Wheeler,J.Opt.Soc.Am.B {\bf {4}},1715(1987).

[16] V.V.Dodonov,A.B.Klimov and V.I.Man'ko,Phys.Lett.A {\bf {134}},211(1989).

[17].V.V.Dodonov,O.V.Man'ko and V.I.Man'ko,Photon distribution for
one-mode mixed
light with generic gaussian Wigner function,University of Napoli
preprint
INFN-NA-IV-93/35,DSF-T-93/35(1993).

[18].V.I.Man'ko, G.Marmo, S.Solimeno, F.Zaccaria, Physical nonlinear aspects
of classical and quantum q-oscillators, Napoli preprint DSF-T-92/25 INFN
-NA-IV-92/25, to appear in Int.Jour.Mod.Phys.;Correlation functions of
quantum q-oscillators,Napoli preprint DSF-T-93/06 INFN-NA-IV-93-06, to
appear in Phys.Lett.A.

\end{document}